\documentstyle[prb,aps]{revtex}
\begin{document}
\draft
\title{Small Angle Scattering by Fractal Aggregates:
      A Numerical Investigation of the Crossover Between
            the Fractal Regime and the Porod Regime}

\author{Anwar Hasmy, Ren\'e Vacher and R\'emi Jullien}
\address{Laboratoire de Science des Mat\'eriaux Vitreux, UA 1119 CNRS,
      Universit\'e Montpellier II, Place Eug\`ene Bataillon,
               34095 Montpellier Cedex 5, France}

\date{\today}
\maketitle

\begin{abstract}
Fractal  aggregates are built on a computer using off-lattice
cluster-cluster aggregation models. The aggregates are made of
spherical  particles of different sizes distributed  according
to a Gaussian-like distribution characterised by a mean $a_0$ and
a  standard  deviation $\sigma$. The wave vector dependent  scattered
intensity $I(q)$ is computed in order to study the influence  of
the  particle  polydispersity on  the  crossover  between  the
fractal  regime and the Porod regime. It is shown that,  given
$a_0$, the location $q_c$ of the crossover decreases as $\sigma$ increases.
The dependence of $q_c$ on $\sigma$ can be understood from the evolution
of  the  shape  of the center-to-center interparticle-distance
distribution function.
\end{abstract}

\pacs{PACS numbers: 61.12.-q, 61.43.Bn, 61.43.Hv}

 Small-angle  X-ray scattering as well as small-angle  neutron
scattering  have  been widely used to study the  structure  of
disordered  systems \cite{1,2,3,4,5,6,7,8}. In particular they have allowed  to
demonstrate that silica aerogels are made of connected fractal
``blobs''\cite{3,4}.  In  such materials, the wave  vector  dependent
scattered  intensity $I(q)$ exhibits two crossovers  related  to
the  two characteristic lengths, mean particle diameter $a_0$ and
mean  blob size $\xi$. The first one separates the low-$q$  ($q<<\xi^{-1}$)
saturation,   called  the  Guinier  regime \cite{9},   from   the
intermediate  power-law behaviour \cite{3,4},  called  the  fractal
regime. The second one, on which we will focus in this letter,
separates  the  fractal regime from the high-$q$  ($q>>a_0^{-1}$) $q^{-4}$
behaviour, called the Porod regime\cite{9}.

 In  some  previous studies, it was assumed that the crossover
between  the  fractal regime and the Porod regime  was  always
exactly located at $q_c=2\pi/a_0$. This assumption was even used  by
some  authors to quantitatively determine the average diameter
\cite{5,6}  or  the  gyration radius \cite{4,7}  of  the  particles.  In
several other papers, an average particle radius was extracted
from  a  fit  of  the scattered intensity $I(q)$  to  analytical
expressions  which do not include polydispersity  \cite{2,8}.  Such
kind  of  analysis has been done in the case of base catalysed
and  neutrally reacted silica aerogels where the particle size
polydispersity impedes the determination of $a_0$ by conventional
methods  such as electron micrography. In this letter we  show
that  such  an  assumption  is only valid  when  the  standard
deviation  $\sigma$  of  the particle size is small compared  to  the
average $a_0$ .

 In  previous  papers \cite{10,11} it has been shown that  cluster-
cluster  aggregation models \cite{12} can satisfactorily  reproduce
the  structure  of  aerogels.  Furthermore,  if  we  are  only
interested  in  short range correlations, it is sufficient  to
consider   a  single  aggregate  obtained  with  a  simplified
aggregation  process:  the hierarchical cluster-cluster  model
\cite{13}.

 Here   for  simplicity  we  have  considered  the   case   of
chemically  limited  cluster-cluster aggregation  (CLCA) \cite{14}
using  a three dimensional off-lattice hierarchical procedure.
The  hierarchical scheme is an iterative method  which  starts
with a collection of $N_p=2^p$ particles at iteration $i=0$ and ends
with a unique aggregate of $N_p$ particles at iteration $p$. At  an
intermediate  iteration  $i$, one has a  collection  of  $N_c=2^{p-i}$
independent   aggregates,  each  of   them   containing  $N=2^i$
particles.  To  proceed  to  the  next  iteration,  the  $2^{p-i}$
aggregates are grouped into pairs and is built a new aggregate
with  each pair according to a specific sticking rule. As soon
as  it  is obtained, the new aggregate is randomly disoriented
and  is  stored in the collection for the next iteration.  The
sticking  rules depend on the chosen aggregation  process.  In
the  CLCA case the sticking rule is as follows: A particle  of
one  cluster and a particle of another cluster are  chosen  at
random  as  well  as  a random direction  in  space.  The  two
clusters  are  disposed such that these two particles  are  in
contact  with their centers aligned along a random  direction.
Then a test of overlap is made for the other particles. If  an
overlap is found, the trial is discarded and another choice is
made  for  both  particles  and random  direction.  Then,  the
resulting aggregate is stored for the next iteration. In  this
model, the resulting fractal dimension \cite{15} is equal to  about
2.

 In  this work the hierarchical procedure is initialised  with
a  collection  of $N$ polydisperse particles whose diameters  $a_i$
are    distributed   according   to   a   truncated   Gaussian
distribution.  We  use  a  standard library  subroutine  which
generates  a  set  of random variables $x_i$  of  zero  mean  and
standard deviation equal to one, distributed according to  the
normal probability law:
\begin{eqnarray} 
g(x)={1\over{\sqrt{2\pi}}}e^{-{x^2\over 2}}
\label{E1}
\end{eqnarray}
The diameters $a_i$ are calculated by:
\begin{eqnarray} 
a_i=1+sx_i\label{E2}
\end{eqnarray}
where  $s$ is an input parameter and where all the $x_i$'s leading
to  negative  $a_i$'s have been discarded. Then we calculate  the
average  $a_0$  and  the  standard deviation  $\sigma$  from  the  usual
formulae:

\begin{mathletters}
\begin{eqnarray}
a_0={1\over N}\sum\limits_{i=1}^N{a_i}\label{E3} 
\end{eqnarray}

{\parindent 0mm \rm and}

\begin{eqnarray}
\sigma^2={1\over N}\sum\limits_{i=1}^N{(a_i-a_0)^2} \label{E4}
\end{eqnarray}
\end{mathletters}

 By  varying  the  parameter $s$, one  can  vary  the  effective
standard  deviation $\sigma_{eff}(=\sigma/a_0)$ which is  the  only  relevant
dimensionless  parameter in our problem. Note  that  $\sigma_{eff}$ is
practically equal to $s$ for $s \leq 0.17$, where $a_0\simeq 1$. In figure  1
we show typical histograms for different values of $\sigma_{eff}$.

 In  the  general  case (where the standard deviation  may  be
large)  one  can  no longer calculate the scattered  intensity
$I(q)$  as  a  product of a form factor $P(q)$ and  the  structure
factor $S(q)$.  One  should go back to the calculation  of  the
scattered amplitude \cite{16},  which is proportional to:
\begin{eqnarray}
\tilde{A}=\sum\limits_i\int\limits_v{e^{i\vec{q}.(\vec{r}_i+\vec{x})}{\rm d}^3x} \label{E5}
\end{eqnarray}
where $\vec{r}_i$ refers to the center of the $i$-th particle and $\vec{x}$  refers
to a running point inside the volume of the $i$-th particle with
respect  to  its  center. The integral inside the  sum,  which
should be performed over the volume of the $i$-th particle,  can
be  calculated  as  a  function of $a_i$, assuming  isotropy  and
homogeneity inside the sphere, leading to:
\begin{mathletters}
\begin{eqnarray}
\tilde{A}=\sum\limits_i{e^{i\vec{q}.\vec{r}_i}A_i(q)} \label{E6}
\end{eqnarray}

{\parindent 0mm \rm with:}

\begin{eqnarray}
A_i(q)=4\pi{\sin({qa_i\over 2})-({qa_i\over 2})\cos({qa_i\over 2})\over {q^3}}\label{E7}
\end{eqnarray}
\end{mathletters}
Then, assuming a random orientation of the aggregate over  the
direction  of $\vec{q}$,  the  scattered intensity  $I(q)=|\tilde{A}|^2$ can  be
written as:
\begin{eqnarray}
I(q)=\sum\limits_{i,j}{A_iA_j{\sin(qr_{ij})\over qr_{ij}}}
\label{E8}
\end{eqnarray}
where $r_{ij}=|\vec{r}_i-\vec{r}_j|$.

 Note  that, since the $i$ and $j$ dependent product $A_i A_j$ appears
inside  the sum, the result cannot be split in two  parts.  In
particular  one cannot use the distance distribution  function $f(r)=\sum\limits_{i,j}\delta(r-r_{ij})$
to  calculate an intermediate structure factor $S(q)$. Here  the
double sum has been calculated directly.

 The  numerical results are reported in figure 2 where we have
plotted the intensity $I(q)$ as a function of the ``reduced'' wave
vector $qa_0$. In figure 2a is shown the $I(q)$ curve obtained  for
$\sigma_{eff}$=0,  in  this  curve  we  can observe  the  three  regimes
mentioned   above.   Figure  2b  shows   three  $I(q)$  curves
corresponding to three aggregates of different $\sigma_{eff}$ values. In
this figure, in order to better see the crossover between  the
fractal  and  the  Porod regime, we use the convenient  $I(q)q^4$
representation  where  the first maximum  corresponds  to  the
crossover.  As indicated by the arrows on the left  of  figure
2b,  the location of the crossover between the Guinier and the
fractal  regime is the same in all cases. Also, the  slope  of
the  fractal regime is unchanged. These two results mean  that
the  particle size polydispersity does not affect  the  intra-
aggregate  long  range particle correlations as  well  as  the
overall  size of the aggregate. However, the location  of  the
maximum corresponding to the crossover between the fractal and
the Porod regime strongly depends on $\sigma_{eff}$. In the monodisperse
case  ($\sigma_{eff}$=0),  one has $q_ca_0=2\pi$, as  expected,  but,  for
increasing $\sigma_{eff}$ values, $q_ca_0$ is pushed towards low values.  On
the   other  hand,  in  figure  2b,  one  observes   how   the
oscillations of the Porod regime are more and more  damped  as
the degree of polydispersity increases.

 In  figure 3 we report the variation of $q_ca_0$ as a function of
$\sigma_{eff}$.  There  is  first  a quite slow decrease,  but  after  a
sigmoidal-like behaviour, $q_ca_0$ reaches, for $\sigma_{eff}>0.1$, a  net
linear behaviour which can be approximated by:
\begin{eqnarray}
q_c={2\pi\over a_0}(1-1.6\sigma_{eff})
\label{E9}
\end{eqnarray}
The  overall  decrease of $q_ca_0$ when increasing polydispersity
can  be  attributed to the fact that larger particles dominate
the scattering.

 In  order  to understand the small change of regime  observed
near $\sigma_{eff}$=0.1 in figure 3, we have calculated the center-to-
center interparticle distribution function $f(r)$ for aggregates
of   different  degrees  of  polydispersity,  which  has  been
normalised as follows:
\begin{eqnarray}
\int\limits_0^{\infty}{f(r)4\pi r^2{\rm d}r}={N-1\over 2}
\label{E10}
\end{eqnarray}
where $N$ is the number of particles of the aggregate.

 In  practice, to calculate $f(r)$, we choose a given  path  $\delta r$,
and  we calculate the number of interparticle distances  lying
between $r$ and $r+\delta r$. Then we divide the result by $4\pi r^2N\delta r$.  In figure  4 we compare $f(r)$ curves
with $\sigma_{eff}$=0.1 and with $\sigma_{eff}=0.16$
with the monodisperse curve ($\sigma_{eff}$=0). The delta peak at
$r/a_0=1$  as well as the discontinuity at $r/a_0=2$,  which  has
been attributed to short range interparticle correlations in a
previous  publication \cite{11}, are progressively washed out  when
introducing  polydispersity. Even if $I(q)$  and $f(r)$ are  not
directly  related, it is worth noticing that the bump observed
in  figure 3 occurs at the value  $\sigma_{eff}$=0.1 at which  the  two
peaks merge into a single broad peak. Therefore the change  of
behaviour  occurring near $\sigma_{eff}$=0.1 can  be  attributed  to  a
modification   of   the  short  range  correlations   due   to
polydispersity. Note that such effects have nothing to do with
the  truncation  of  the Gaussian distribution  that  we  have
considered since this truncation becomes effective for $\sigma_{eff}>
0.17$.  On the other hand, we can understand why the $I(q)$ curve
presents only one maximum at high-$q$ for $\sigma_{eff}>0.1$ (see  figure
2b).

 We  have  considered other aggregation models: the  diffusion
limited   cluster-cluster  aggregation  (DLCA) \cite{17,18}   and
chemically limited particle-cluster aggregation (the so-called
Eden model \cite{19}). The first model gives fractal aggregate with
a fractal dimension $D\simeq 1.78$, and the second model corresponds
to  a homogeneous aggregate of dimension equal to 3. In figure
5  we show the two $I(q)$ resulting curves for a fixed value  of
$\sigma_{eff}$, in comparison with the CLCA case. In this figure, we see
that the crossover between the fractal regime and Porod regime
remains at the same $q_ca_0$ value. This fact is due to the common
short  range correlations in all the considered cases, meaning
that  relation  (7)  is  valid for  a  wide  range  aggregated
systems.

 In  summary, we have demonstrated that the crossover  between
the  fractal regime and the Porod regime is very sensitive  to
the particle polydispersity. An empirical relation between the
crossover  wave vector $q_c$, the particle average  diameter  $a_0$
and  the standard deviation $\sigma$ was obtained. We have also shown
that  the  results are almost independent on  the  aggregation
mechanism.

 We  acknowledge interesting discussions with M. Foret. One of
us  (A.H.)  would  like  to acknowledge support  from  CONICIT
(Venezuela).

     % References

   % FIGURE CAPTIONS

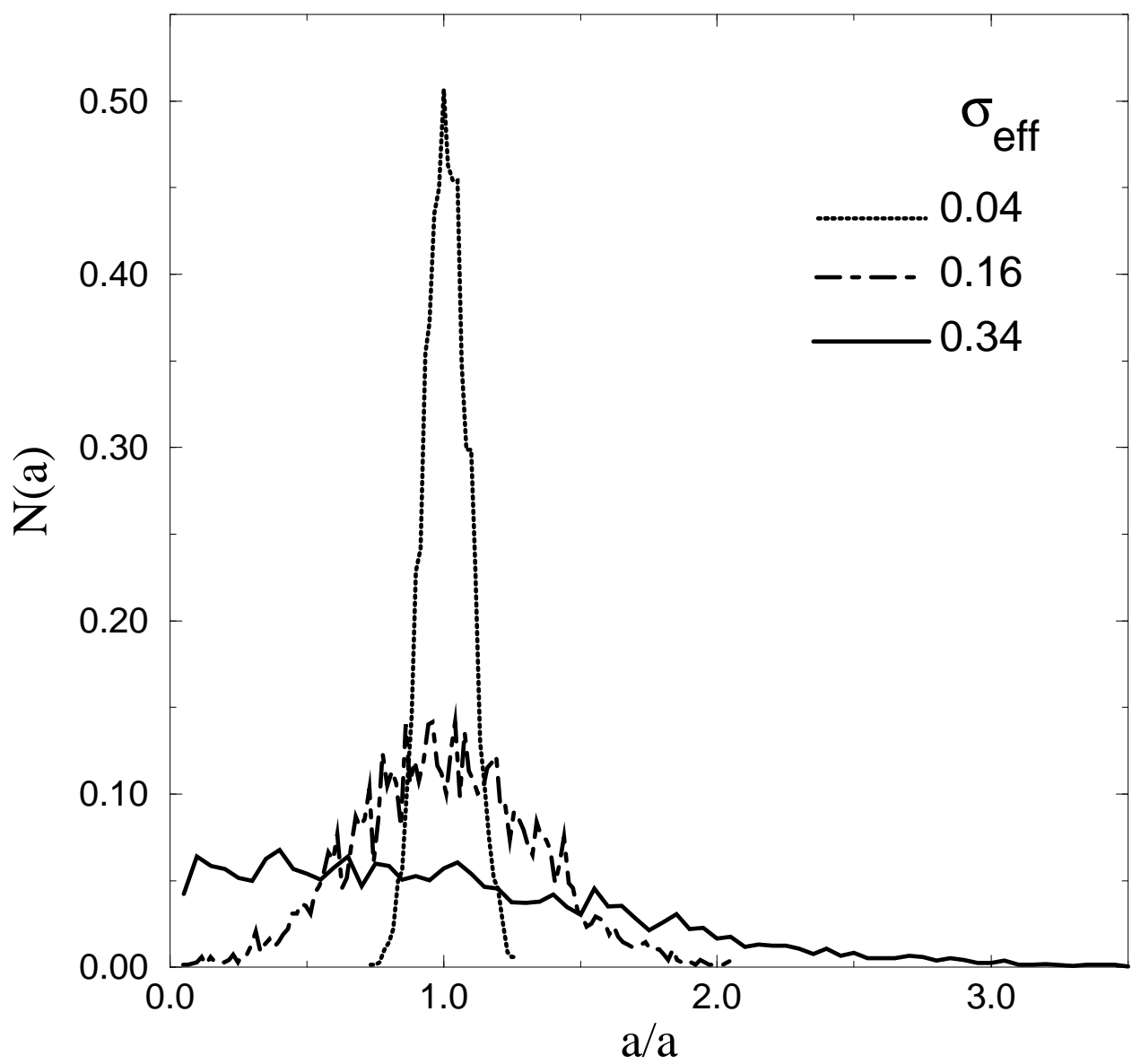
\begin{figure} 
\caption{Histograms  of  the particle  size  Gaussian-like distribution for three different $\sigma_{eff}$ values.}\end{figure}

\begin{figure}
\caption{
(a) Log-log plot of $I(q)$ versus $qa_0$ for  $\sigma_{eff}$=0.
(b) Log-log plot of $I(q)q^4$ versus $qa_0$ for three different $\sigma_{eff}$ \
values.  The  white arrows indicate the crossover between  the
Guinier and the fractal regime. The black arrows indicate  the
crossover between the fractal and the Porod regime. All  these
curves  result from an average over 32 simulations with  $N=128$
particles.
}
\end{figure}

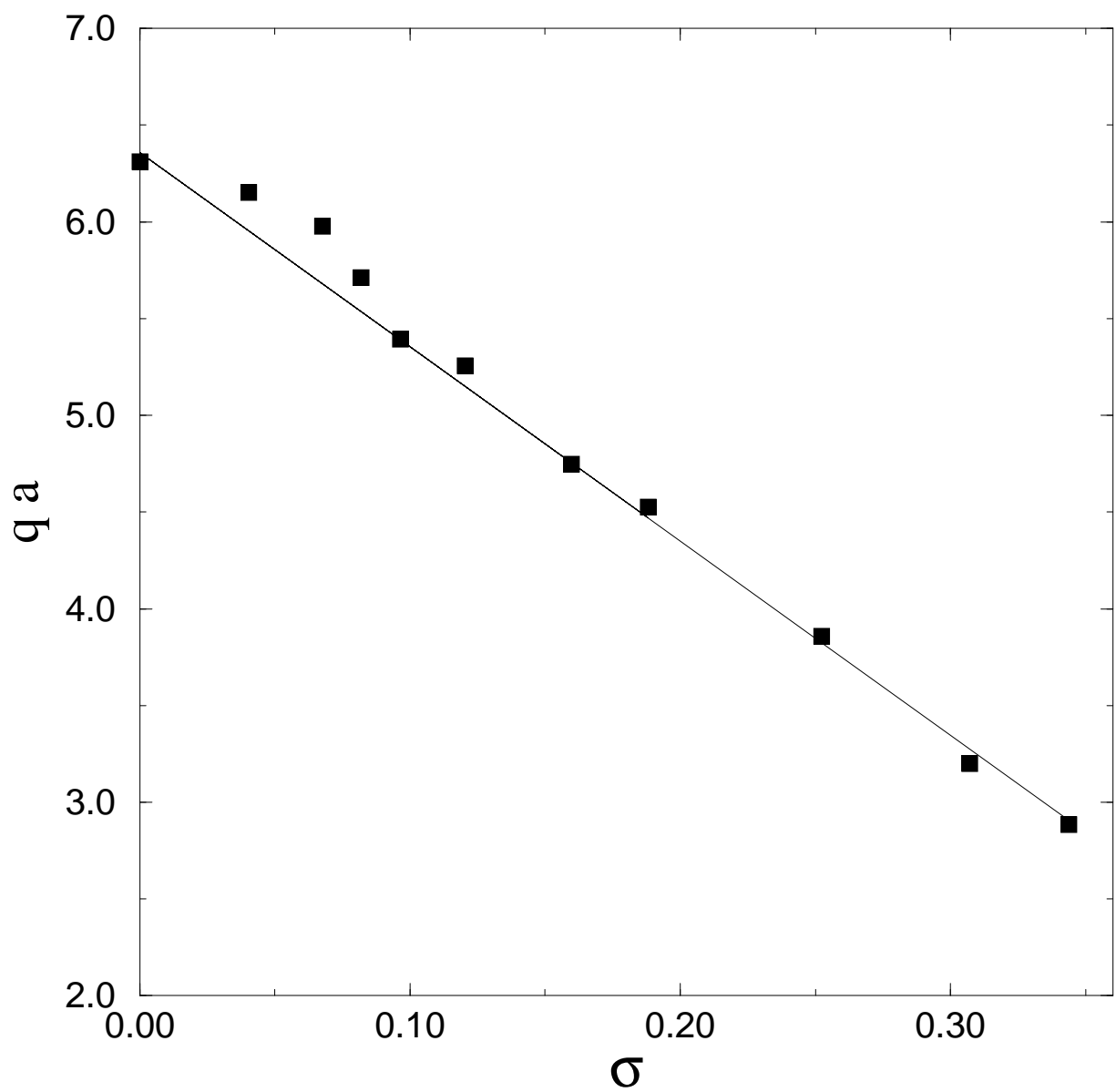
\begin{figure}
\caption{Plot of $q_ca_0$ versus $\sigma_{eff}$.}\end{figure}

\begin{figure}
\caption{
Plots of $f(r)$ versus $r$ for three $\sigma_{eff}$ values.  All
these  curves result from an average over 32 simulations  with
$N=128$ particles.
}
\end{figure}
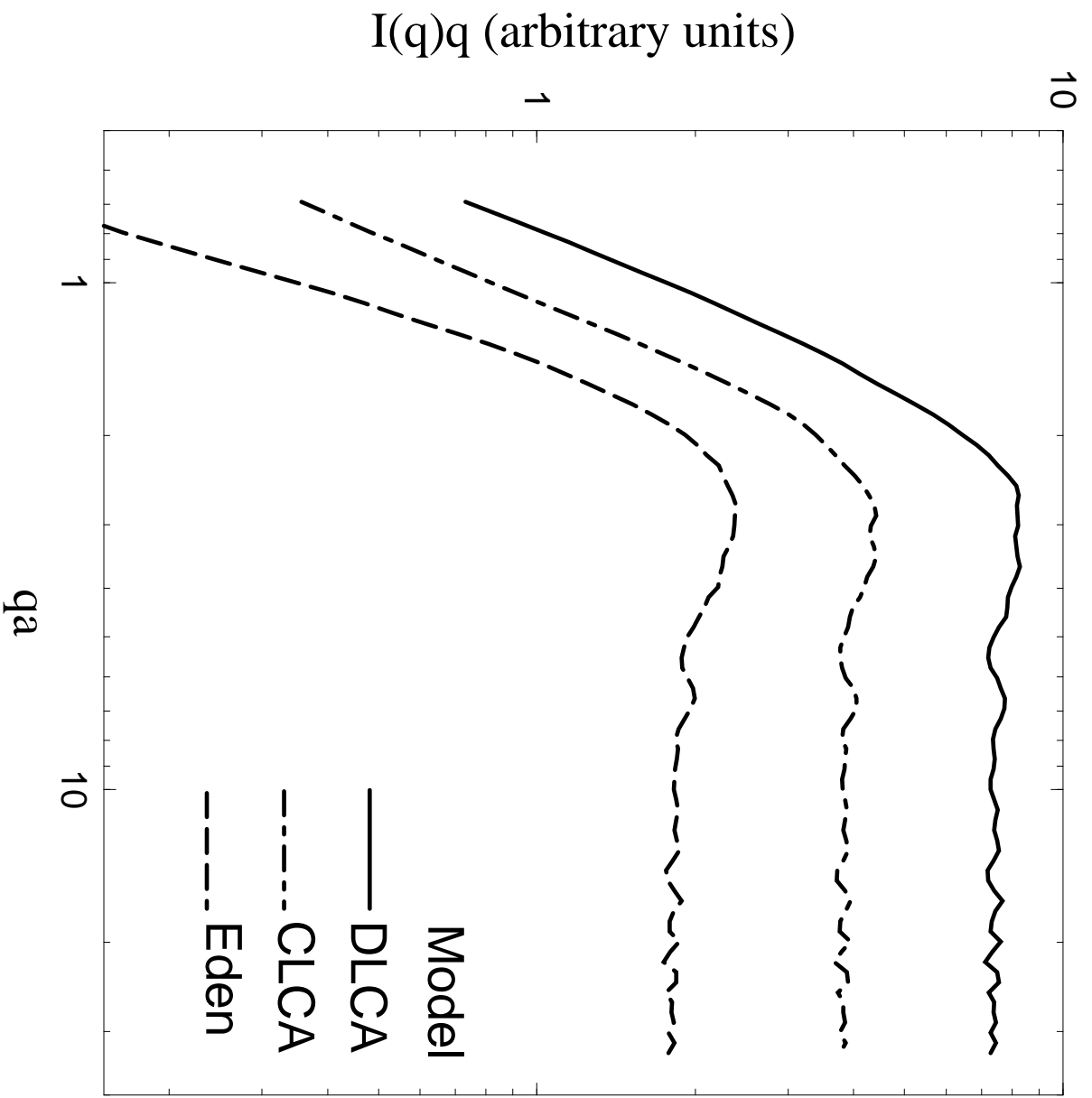
\begin{figure}
\caption{
Log-log  plots of $I(q)q^4$  versus  $qa_0$  for  three
different  models  with $\sigma_{eff}$=0.34. All these  curves  result
from an average over 32 simulations with $N=128$ particles.
}
\end{figure}


\begin{references} 
\bibitem{1}  S. H. Chen and J. Teixeira, Phys. Rev. Letters, {\bf 57},
2583 (1986)                                            
\bibitem{2}  T. Freltoft, J. K. Kjems, and S. K. Sinha, Phys Rev.  B,
{\bf 33}, 269 (1986)
\bibitem{3}  G.  Dietler,  C. Aubert, and D. S. Cannell,  Phys.  Rev.
Letters, {\bf 57}, 3117 (1986)
\bibitem{4}  R. Vacher, T. Woigner, J. Pelous, and E. Courtens, Phys.
Rev. B, {\bf 37}, 6500 (1988)
\bibitem{5}  D. W. Schaefer and K. D. Keefer, Phys. Rev. Letters, {\bf 56},
2199 (1986)
\bibitem{6}  A. Boukenter, D. Champagnon, J. Dumas, E. Duval,  J.  F.
Quinson, J. L. Rousset, J. Serughetti, S. Etienne, and C. Mai,
Revue Phys. Appl. (Paris), {\bf 24}, C4-133 (1989)
\bibitem{7}  F.  Chaput,  A. Lecomte, A. Dauger, and  J.  P.  Boilot,
Revue Phys. Appl. (Paris), {\bf 24}, C4-137 (1989)
\bibitem{8}  D. Posselt, J. S. Pedersen, and K. Mortensen, J. of Non-
Cryst. Solids, {\bf 145}, 128 (1992)
\bibitem{9} A. Guinier and J. Fournet, {\it Small Angle Scattering of  X-
rays}, (Wiley Interscience, New York, 1955)
\bibitem{10}  A.  Hasmy,  E. Anglaret, M. Foret, J.  Pelous,  and  R.
Jullien (pre-print)
\bibitem{11}  A.  Hasmy,  M. Foret, J. Pelous, and R. Jullien,  Phys.
Rev. B, {\bf 48}, 9345 (1993)
\bibitem{12}  R.  Jullien  and  R.  Botet, {\it Aggregation  and  Fractal
Aggregates}, (World Scientific, Singapore, 1987)
\bibitem{13}  R. Botet, R. Jullien, and M. Kolb, J. Phys. A, {\bf 17},  L75
(1983)
\bibitem{14} R. Jullien and M. Kolb, J. Phys. A, {\bf 17}, L639 (1984)
\bibitem{15}   B.   B.   Mandelbrot,  {\it Fractals:  Form  Chance,   and
Dimension} (Freeman, San Francisco, 1977)
\bibitem{16}  L. A. Feigin and P. I. Svergun, {\it Structure Analysis  by
Small Angle X-rays and Neutron Scattering}, (Plenum, New  York
and London, 1987)
\bibitem{17} P. Meakin, Phys. Rev. Letters, {\bf 51}, 1119 (1983)
\bibitem{18}  M.  Kolb, R. Botet, and R. Jullien, Phys. Rev. Letters,
{\bf 51}, 1123 (1983)
\bibitem{19}  M.  Eden, {\it Proc. 4th Berkeley Symp. Math. Stat. Prob.},
{\bf 4}, 223 (Berkeley: University of California Press, 1961)
\end{references}
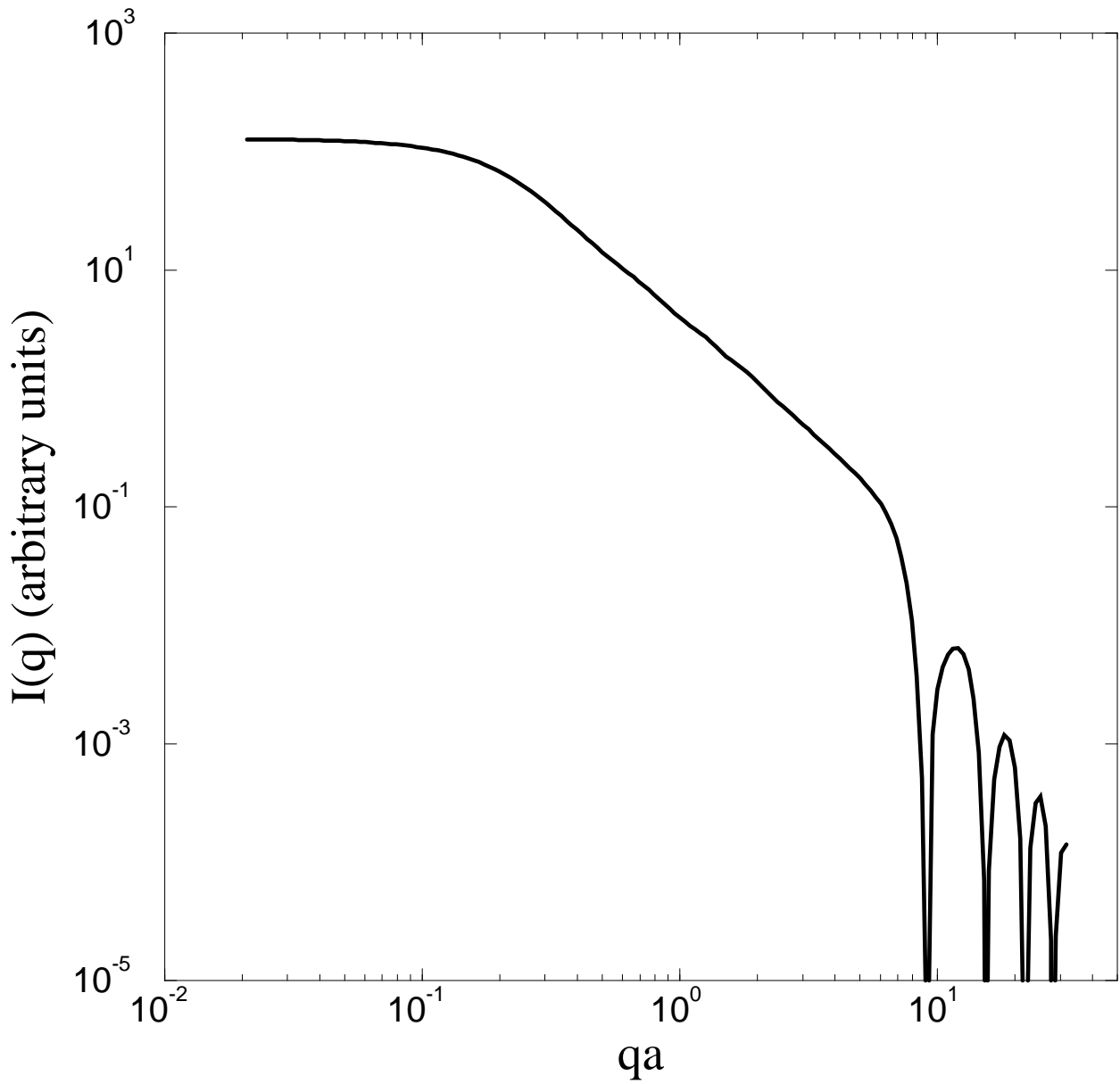
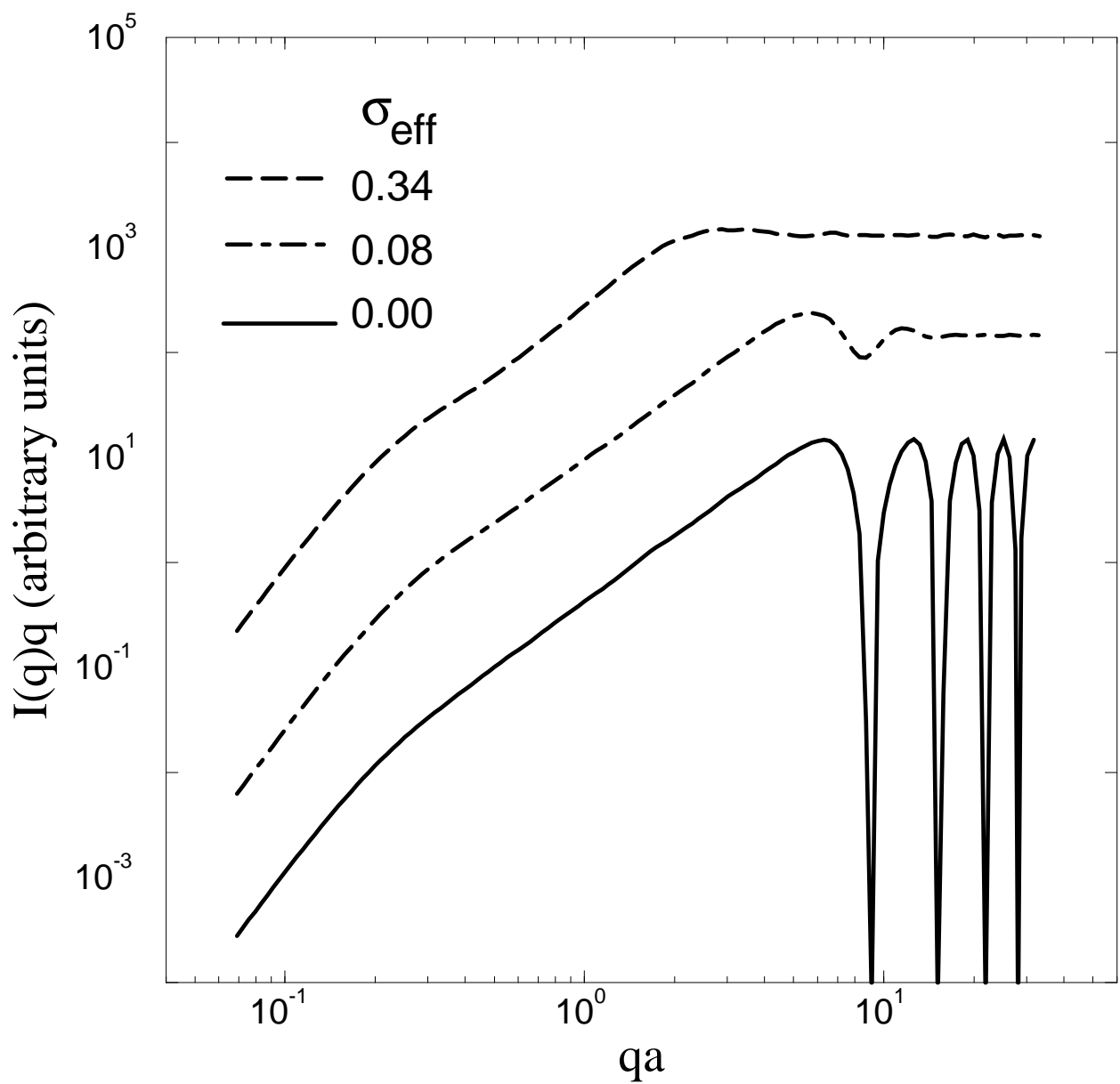
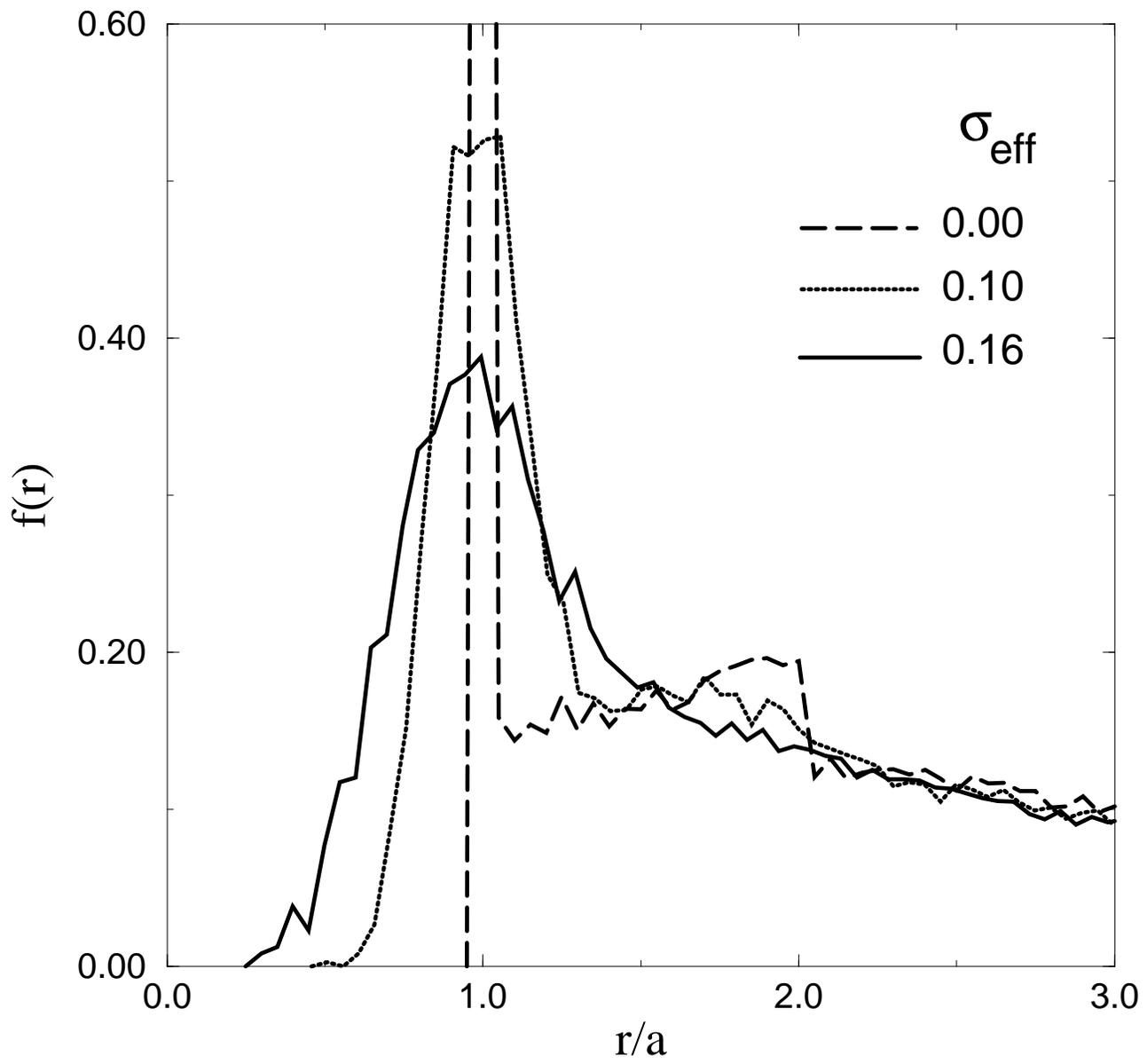
\end{document}